\newcommand{\rem}[1]{}
\documentclass{amsart}
\usepackage{amsfonts,amssymb,amsmath,amsthm,mathrsfs}
\usepackage{url}
\usepackage[dvips]{epsfig}
\newtheorem{thrm}{Theorem}[section]

\newtheorem{prop}[thrm]{Proposition}
\newtheorem{cor}[thrm]{Corollary}
\newtheorem{remark}[thrm]{Remark}
\theoremstyle{definition}
\newtheorem{definition}[thrm]{Definition}

\begin{document}

\author[C.~A.~Mantica and L.~G.~Molinari]
{Carlo~Alberto~Mantica and Luca~Guido~Molinari}
\address{C.~A.~Mantica: I.I.S. Lagrange, Via L. Modignani 65, 
20161, Milano, Italy -- L.~G.~Molinari (corresponding author): 
Physics Department,
Universit\'a degli Studi di Milano and I.N.F.N. sez. Milano,
Via Celoria 16, 20133 Milano, Italy.}
\email{carloalberto.mantica@libero.it, luca.molinari@mi.infn.it}
\title[Weyl compatible tensors]{Weyl compatible tensors}

\begin{abstract}
We introduce the new algebraic property of Weyl compatibility for 
symmetric tensors and vectors. It is strictly related to
Riemann compatibility, which generalizes the Codazzi condition while 
preserving much of its geometric implications. 
In particular it is shown that 
the existence of a Weyl compatible vector implies the Weyl tensor to be 
algebraically special, and it is a necessary and sufficient condition for 
the magnetic part to vanish. 
Some theorems (Derdzi\'nski and 
Shen, Hall) are extended to the broader hypothesis of Weyl or Riemann 
compatibility; Weyl compatibility includes conditions that were
investigated in the literature of general relativity (as McIntosh et al.) 
Hypersurfaces of pseudo Euclidean spaces provide a simple 
example of Weyl compatible Ricci tensor.
\end{abstract}
\date{18 jan 2013}
\maketitle
\section{Introduction}
The geometry of Riemannian or pseudo-Riemannian manifolds 
of dimension $n\ge 3$ is intrinsically described by
$ \mathscr N =\frac{1}{12}n(n-1)(n-2)(n+3) $  algebraically independent 
scalar fields, constructed with the Riemann and the metric tensors. 
The same counting is provided by the Weyl, the Ricci and the metric tensors.
The Weyl tensor bears the symmetries of the Riemann tensor, with the extra 
property of being traceless\footnote{Conventions: 
$X_{[ab]}=: X_{ab}-X_{ba}$, $R_{ab}=R_{amb}{}^m$, $u^2=u^au_a$.}:
\begin{eqnarray}
\qquad C_{jkl}{}^m = R_{jkl}{}^m \,+\, \frac{1}{n-2}
(\delta_{[j}{}^m R_{k]l} +R_{[j}{}^m g_{k]l})  
-\frac{1}{(n-1)(n-2)}\, R\delta_{[j}{}^m g_{k]l}.\label{Weyltensor}
\end{eqnarray}
The trace condition  $C_{jab}{}^j=0$ reduces the parameters of
the Riemann tensor by a number $\frac{1}{2}n(n+1)$ that is accounted 
for by considering the Ricci tensor as algebraically independent.
The two tensors are linked by functional relations 
as the following one \cite{CM2011,HawkingEllis}:
\begin{align}
- \nabla_m C_{abc}{}^m = \frac{n-3}{n-2}\left[\nabla_{[a} R_{b]c} -
\frac{1}{2(n-1)} \nabla_{[a} g_{b]c}R \right ].\label{funct}
\end{align} 
In the coordinate frame that locally diagonalizes the Ricci and the metric 
tensors (the latter with diagonal elements $\pm 1$), the parameters that 
survive are the components of the Weyl tensor and the $n$ eigenvalues 
of the Ricci tensor, whose number is precisely $\mathscr N$ \cite{Weinberg}.
This choice of fundamental tensors offers advantages, as in the classification 
of manifolds and in general relativity.

Debever and Penrose \cite{Penrose} proved that in four-dimensional 
space-time manifolds the equation 
\begin{align}
 k_{[b}C_{a]rs[q}k_{n]}k^r k^s = 0\label{333} 
\end{align} 
always admits four null solutions (principal null directions). When two or 
more coincide, the Weyl tensor is named {\em algebraically special}, and the 
condition for degeneracy is
\begin{equation}
k_{[b}C_{a]rsq}k^r k^s = 0  \label{4}
\end{equation}
The degeneracies classify space-time manifolds in classes that coincide with 
the Petrov types, which are determined by the degeneracies of the eigenvalues 
of the self-dual part of the Weyl tensor \cite{Petrov}.\\ 
For $n>4$, Milson et al. showed in 2005 that eq.~\eqref{333}
may not have solution at all \cite{Milson}. 
They introduced the notion of {\em Weyl aligned null directions} (WAND): 
a null vector $k$ is a WAND 
if there is a null frame including it, such that the Weyl scalars of 
maximal boost weight vanish. 
This is true if $C_{0i0j}=0$. For any $n$ the 
condition is equivalent to $k$ being a solution of eq.\eqref{333} 
(\cite{Milson} Prop.~IV.5). The order of
alignment provides the backbone of a classification of 
Lorentzian manifolds (\cite{Milson} Table III, and 
\cite{Ortaggio}).

The Einstein equations of general relativity link the energy 
momentum tensor $T_{ij}$ to the Ricci tensor and the curvature scalar, but 
not to the Weyl tensor:
\begin{equation}
R_{ij}-\frac{R}{2}\, g_{ij}=8\pi\,T_{ij}. \label{EE}
\end{equation} 
\noindent
In $n=4$ the Weyl tensor may be replaced by two symmetric tensors, the electric
and magnetic components, and the identity \eqref{funct} for the Weyl tensor 
translates into Maxwell-like equations for the components 
\cite{Stephani2, Berts}. The 
construction was extended to $n>4$ \cite{Ortaggio2}.

In the study of Derdzi\'nski and Shen's theorem \cite{DerShen,Besse} 
on the restrictions imposed by a Codazzi tensor on the structure of 
the Riemann tensor, we introduced the new algebraic notion of {\em Riemann 
compatible tensors} \cite{manticaDS}. This enabled us to extend the 
theorem in two directions: the replacement of the Codazzi condition with the 
milder hypothesis of Riemann compatibility together with a drastic 
simplification of the proof, the restatement of the theorem for curvature 
tensors other than Riemann's.
Riemann compatible tensors were investigated in \cite{RCT}. Most of the 
statements valid for (pseudo) Riemannian manifolds equipped with a nontrivial 
Codazzi tensor, such as the vanishing of Pontryagin forms, were shown to 
persist in presence of a nontrivial Riemann compatible tensor. The 
application to geodesic mappings was then discussed.\\ 

This paper is mainly about Weyl compatibility of symmetric tensors, a 
property which is broader than Riemann compatibility. 
The restriction on the structure of the Weyl tensor has consequences on the
Petrov type of the manifold and the electric and magnetic components of 
the Weyl tensor. Since classifications of Lorentzian manifolds are mainly based
on vectors, 
our discussion of Weyl compatibility of vectors crosses in several points 
definitions and properties proven by other authors, but having a 
different origin.

Definitions and main properties of Riemann and Weyl compatibile symmetric 
tensors are reviewed in Sect.~2, with the subcase of Riemann 
and Weyl-{\em permutable} tensors. Tensors $u_iu_j$ naturally 
define Riemann and Weyl compatibility for vectors, which is discussed in 
Sect.~3 with various new results, as the extension of Derdzi\'nski and 
Shen's theorem \cite{DerShen} and of Hall's theorem \cite{Hall}, and
a sufficient condition for the vanishing of Pontryagin forms.  
Riemann (Weyl) 
permutable vectors are considered in Sect.~4, where results by McIntosh and
others \cite{McIvLee,McIH,McIHa}
for the special case $R_{ijkl}u^l=0$ are reobtained and extended.\\
In Sect.~5 it is shown that the existence of a Weyl compatible vector is a 
sufficient condition for the Weyl tensor to be special, with
results regarding the Penrose-Debever classification of spacetimes.
The electric and magnetic components of the Weyl tensor
are considered 
in Sect.~6, with the statements that existence of a Weyl compatible vector is 
necessary and sufficient for the Weyl tensor to be purely electric, and 
that Weyl permutability implies a conformally flat space-time.\\
Sect.~7 is devoted to hypersurfaces; the Gauss and Codazzi equations 
specify the induced Riemann tensor of the hypersurface as a quadratic
expression of a Codazzi tensor. It is shown that the corresponding Ricci 
tensor is Riemann and Weyl compatible.\\
Conformal maps obviously preserve Weyl compatibility, geodesic maps 
preserve Riemann compatibility \cite{RCT} but not necessarily Weyl 
compatibility; a sufficient condition is presented in Sect.~8.

The manifolds considered here are Hausdorff connected with non
degenerate metric of arbitrary signature, i.e. $n$-dimensional 
pseudo-Riemannian manifolds. Where necessary, we specialize to
the metric signature $n-2$, i.e. to $n$-dimensional Lorentzian 
manifolds (space-times).
We always assume a Levi-Civita connection ($\nabla_i g_{jk}=0 $).

\section{Riemann and Weyl compatible tensors}
We briefly review the concept of compatibility for symmetric tensors, first
introduced in \cite{manticaDS}, and investigated in \cite{RCT}. Permutable 
tensors are then defined, as a special class.  
\begin{definition}
A symmetric tensor $b_{ij}$ is Riemann compatible if:
\begin{equation}
b_{am} R_{bcl}{}^m + b_{bm} R_{cal}{}^m + b_{cm} R_{abl}{}^m = 0.
\label{Rcomp}
\end{equation}
\end{definition}
\noindent
The metric tensor is trivially Riemann compatible because of the first Bianchi
identity.

\begin{remark}
The definition has a natural origin. Consider the vector-valued 1-form 
$B_l = b_{kl} dx^k $, where $b$ is a symmetric tensor. A covariant exterior
derivative gives $$ D B_l=\frac{1}{2} \mathscr C_{jkl}\, dx^j \wedge dx^k, $$ 
where $\mathscr C_{ijk} =: \nabla_i b_{jk}-\nabla_jb_{ik} $ is the 
``Codazzi deviation tensor'', defined in \cite{RCT}.
As it is well known, $DB_l=0$ if and only if $b_{kl}$ is a Codazzi 
tensor \cite{Bourguignon,Suh}.\\
If $DB_l\neq 0$, another derivative gives 
\begin{equation*}
D^2B_l=
\frac{1}{3!}\left (
\nabla_i\mathscr C_{jkl}+\nabla_j\mathscr C_{kil}+\nabla_k\mathscr C_{ijl}
\right )
dx^i\wedge dx^j\wedge dx^k. 
\end{equation*}
The following identity links the Codazzi deviation to Riemann 
compatibility \cite{RCT}:
\begin{align}
\nabla_i\mathscr C_{jkl}+\nabla_j\mathscr C_{kil}+\nabla_k\mathscr C_{ijl} =
b_{im} R_{jkl}{}^m+b_{jm} R_{kil}{}^m+b_{km} R_{ijl}{}^m\label{CCCcomp}
\end{align}
Therefore, $D^2B_l=0 $ (i.e. $DB_l$ is closed) if and only if $b$ is 
Riemann compatible.\\
It follows that Codazzi tensors, $\nabla_i b_{jk}=\nabla_j b_{ik}$, 
are Riemann compatible.
\end{remark}
\noindent
As an example, consider the Ricci tensor. Its Codazzi deviation is
$\mathscr C_{abc}=: \nabla_{[a}R_{b]c} = -\nabla_m R_{abc}{}^m$, by the
contracted Bianchi identity. The identity \eqref{CCCcomp} turns out to 
be Lovelock's identity \cite{Lovelock}:
\begin{align}
-(\nabla_a\nabla_m R_{bcd}{}^m+\nabla_b\nabla_m R_{cad}{}^m+
\nabla_c\nabla_m R_{abd}{}^m )= \label{Lovelock}  \\
R_{am}R_{bcd}{}^m+R_{bm}R_{cad}{}^m+R_{cm}R_{abd}{}^m \nonumber
\end{align}

Compatibility was extended to generalized curvature tensors $K_{abc}{}^m$, 
i.e. tensors having the symmetries of 
the Riemann tensor in the exchange of indices, 
and the first Bianchi property. 
The Weyl tensor \eqref{Weyltensor} is the most notable example. 
A symmetric tensor $b_{ij}$ is {\em Weyl compatible} if:
\begin{equation}
b_{am} C_{bcl}{}^m + b_{bm} C_{cal}{}^m + b_{cm} C_{abl}{}^m = 0.
\label{Wcomp}
\end{equation}
In \cite{CM2011} (Prop.~2.4) we showed an invariance property of Lovelock's 
identity \eqref{Lovelock}. In particular, it remains valid if the Riemann 
tensor in the left hand side is replaced by the Weyl tensor. If Einstein's
equations are then used, one gets a differential condition for the Weyl 
tensor involving the stress-energy tensor $T_{ij}$:
\begin{align*}
& \nabla_i\nabla_m C_{jkl}{}^m  +\nabla_j\nabla_m C_{kil}{}^m
+ \nabla_k\nabla_m C_{ijl}{}^m  \\
&=-8\pi \frac{n-3}{n-2}\left( T_{im}C_{jkl}{}^m +T_{jm}C_{kil}{}^m + 
T_{km}C_{ijl}{}^m   \right)
\end{align*}
Since the left hand side is the exterior covariant differential of the vector 
valued 1-form 
$\Pi_l = \nabla_m C_{jkl}{}^m dx^j\wedge dx^k$, the following theorem 
holds:
\begin{thrm}\label{DPi}
On a $n$-dimensional pseudo-Riemannian manifold, the Ricci tensor and
the energy-stress tensor are Weyl compatible if and only if $D\Pi_l=0$.
\end{thrm}
\noindent
The condition $D\Pi_l=0$ is satisfied in $n$-dimensional 
Lorentzian manifolds that are conformally symmetric 
($\nabla_i C_{jkl}{}^m=0$) or conformally recurrent 
($\nabla_i C_{jkl}{}^m = \alpha_i C_{jkl}{}^m$). On these manifolds  
the Ricci and the stress energy tensors are  Weyl compatible \cite{CM2011,Suh}.

A Weyl compatible tensor poses strong restrictions on the Weyl tensor. 
In \cite{manticaDS} we proved a broad generalization of Derdzinski and 
Shen's theorem that holds both in Riemannian and pseudo-Riemannian manifolds.
For the Weyl tensor it reads:
\begin{prop}\label{CXYZ}
On a pseudo-Riemannian manifold with a Weyl compatible tensor $b$, 
if $X$, $Y$ and $Z$ are eigenvectors of $b$ with eigenvalues 
$\lambda $, $\mu $, $\nu $ ($b^i{}_jX^j=\lambda X^i$, etc.) then:
\begin{equation}
C_{abcd}X^aY^bZ^c = 0 , \quad \nu\neq\lambda , \mu. 
\end{equation}
\end{prop}
{\quad}\\
The following algebraic identity relates a symmetric tensor $b_{ij}$ to 
the Weyl, the Riemann and the Ricci tensors \cite{RCT}:
\begin{align}
b_{im} C_{jkl}{}^m + b_{jm} C_{kil}{}^m + b_{km} C_{ijl}{}^m 
= b_{im} R_{jkl}{}^m + b_{jm} R_{kil}{}^m + b_{km} R_{ijl}{}^m \label{RC}\\
+ \frac{1}{n-2}\left [ 
g_{kl} (b_{im} R_j{}^m - b_{jm} R_i{}^m ) + g_{il} (b_{jm} R_k{}^m - b_{km} 
R_j{}^m ) + g_{jl} (b_{km} R_i{}^m - b_{im} R_k{}^m )\right ].\nonumber
\end{align}
Any contraction with the metric tensor gives zero; the identity is 
trivial if $b$ is the metric tensor. An immediate consequence is:
\begin{thrm}\label{bRWRicci}
A symmetric tensor is Riemann compatible if and only if 
it is Weyl compatible and it commutes with the Ricci tensor.
\begin{proof}
If $b$ is Riemann compatible, contraction of \eqref{Rcomp} with $g^{cl}$
gives $b_{am} R_b{}^m - b_{bm} R_a{}^m =0$, i.e. $b$ commutes with the Ricci
tensor. Then $b$ is Weyl compatible by identity \eqref{RC}. The converse
is obvious, by the same identity.
\end{proof}
\end{thrm}
\noindent
In particular, Riemann and Weyl compatibility are equivalent for the Ricci
tensor, or any symmetric tensor that commutes with it.\\

An example of Riemann tensor with a Riemann compatible symmetric
tensor can be constructed by the Kulkarni-Nomizu product of two
symmetric tensors \cite{Besse,Bourguignon,DerShen}: 
\begin{prop}
Suppose that the Riemann tensor has the form:
\begin{equation}\label{RBA}
R_{jklm}= b_{l[j}a_{k]m}+b_{m[k}a_{j]l} 
\end{equation}
with symmetric tensor fields $a_{ij}$ and $b_{ij}$. If they
commute, $a_i{}^mb_{mj}= a_j{}^mb_{mi}$, then they are both 
Riemann compatible.
\begin{proof}
Evaluate: $b_i{}^mR_{jklm}=b_{lj}(ba)_{ik}-b_{lk}(ba)_{ij}+b^2_{ik}a_{jl}-
b^2_{ij}a_{kl}$. The sum on cyclic permutations of $ijk$ cancels  
the r.h.s. i.e. $b$ is Riemann compatible. Because of the symmetry of 
\eqref{RBA} in the exchange of $a$ and $b$, also $a$ is Riemann compatible. 
\end{proof}
\end{prop}
\noindent
The commuting tensors $a,b$ are also Weyl compatible for the   
Weyl tensor computed from \eqref{RBA}. The same Kulkarni-Nomizu product can 
be used to construct another Weyl tensor:
\begin{prop}
Let $a$ and $b$ be commuting symmetric tensor fields, such that  
\begin{equation}\label{condition}
b^m{}_m a_{kl}+a^m{}_m b_{kl}-2 b_{km}a^m{}_l=0,
\end{equation} 
then $C_{jklm}= b_{l[j}a_{k]m}+b_{m[k}a_{j]l} $
has the symmetries of the Weyl tensor, and $a$ and $b$ are Weyl compatible.
\end{prop}
\noindent
The additional equation \eqref{condition} is required 
to enforce tracelessness, and can be solved to obtain the ``potential'' $a$
that produces $b$.
%
\\

Suppose that a symmetric tensor has the property
$b_{im} R_{jkl}{}^m = \omega \,b_{lm} R_{jki}{}^m  $, where $\omega $ is
a scalar. Then either $\omega =\pm 1 $ or $b_{im} R_{jkl}{}^m=0$.
The three cases define interesting classes of tensors that will be
shown to be Riemann compatible. The same conditions are found with
the Weyl tensor.\\
The class $\omega =-1$ was studied by McIntosh and others
\cite{McIHa,McIH} and is presented in Sect.~4. 
The class $b_{im} R_{jkl}{}^m=0$ was studied
in \cite{McIvLee} for $b_{ij}=u_iu_j$, i.e $R_{jkl}{}^m u_m=0$.
Now we consider the class $\omega =1$: 

\begin{definition}
A symmetric tensor $b_{ij}$ is {\em Riemann permutable} if:
\begin{align}
b_{im} R_{jkl}{}^m = b_{lm} R_{jki}{}^m \label{permut}
\end{align} 
It is {\em Weyl permutable} if:
\begin{align}
b_{im} C_{jkl}{}^m = b_{lm} C_{jki}{}^m \label{Wpermut}
\end{align}
\end{definition}
\begin{prop}
If a symmetric tensor is Riemann (Weyl) permutable then it is Riemann 
(Weyl) compatible.
\begin{proof}
In the relation for Riemann compatibility use \eqref{permut} for each 
term:
$b_{im} R_{jkl}{}^m + b_{jm} R_{kil}{}^m + b_{km} R_{ijl}{}^m = b_{lm} 
( R_{jki} {}^m + R_{kij}{}^m + R_{ijk}{}^m ) = 0$ by the first Bianchi 
identity. An analogous proof holds for Weyl permutable tensors.
\end{proof}
\end{prop}
\noindent
Note that Riemann permutability does not imply Weyl permutability.  
A Riemann permutable tensor (being Riemann compatible) commutes with 
the Ricci tensor.

Derdzi\'nski and Shen's theorem for the Riemann tensor and theorem
\ref{CXYZ} for the Weyl tensor, become more stringent for permutable 
tensors:
\begin{prop}\label{permXY}
If $b$ is a symmetric tensor and $X$, $Y$ are two eigenvectors, 
$b^i{}_j X^j=\lambda X^i$ and $b^i{}_jY^j=\mu Y^i$ with $\lambda+\mu\neq 0$, 
then\\
1) if $b$ is Riemann permutable it is: $ R_{jklm}X^lY^m = 0$;\\
2) if $b$ is Weyl permutable it is: $ C_{jklm}X^lY^m = 0$. 
\begin{proof}
Contraction of \eqref{permut} with $X^iY^l$ gives
$\lambda R_{jkl}{}^m X_mY^l= \mu R_{jki}{}^m X^iY_m$, then
$0=(\lambda+\mu ) R_{jklm}X^lY^m$. The proof for the Weyl tensor is 
analogous.
\end{proof}
\end{prop}

\section{Riemann and Weyl compatible vectors}
The notion of $K$-compatible symmetric tensor includes 
vectors $u_i$ in a natural way, through the symmetric tensor $u_iu_j$:
\begin{definition} 
A vector field $u_i$ is 
$K$-compatible (where $K$ is the Riemann, the Weyl or a generalized tensor) if:
\begin{align}
(u_i K_{jkl}{}^m + u_j K_{kil}{}^m + u_k K_{ijl}{}^m) u_m =0 \label{uKcomp}
\end{align}
\end{definition}

On a Lorentzian manifold, let $K$ be the Riemann (Weyl) tensor. If $u$ is a
time-like vector, the definition corresponds to the statement that $u_iu_j$
is a {\em purely electric} Riemann (Weyl) tensor (\cite{Ortaggio2} Prop.~3.5
and Prop.~4.3). If $u$ is a null vector, it is a double WAND and the 
manifold is type II(d) (\cite{Ortaggio}, Table~1).

\begin{thrm}
A vector field $u$ with $u^2\neq 0$ is $K$-compatible if and only if
there is a symmetric tensor $D_{ij}$ such that:
\begin{align}
K_{abcm} u^m = D_{ac}u_b - D_{bc}u_a \label{KDD}
\end{align} 
\begin{proof}
Multiplication by $u_d$ and cyclic summation on $abd$ makes the r.h.s. vanish
and $K$-compatibility is obtained.\\
If $u$ is $K$-compatible then multiplication of \eqref{uKcomp} by $u^i$ gives
\begin{align*}
(u^2 K_{jkl}{}^m - u_j u^i K_{ikl}{}^m + u_k u^i K_{ijl}{}^m) u_m =0 
\end{align*}
where we read $D_{jl} = K_{ijlm}u^i u^m/u^2$.
\end{proof}
\end{thrm}
\noindent
It can be easily shown that $u$ is an eigenvector of the symmetric
tensor $D$. For the Weyl tensor, $D$ will be identified with its electric 
component (see Sect.~5).\\ 
On a Lorentzian manifold, the theorem with $K=C$ (Weyl tensor) follows 
from eq.~20 in \cite{Ortaggio2}.

\begin{remark}\label{concircular}
Suppose that on a pseudo-Riemannian manifold there is a concircular vector,
$\nabla_k u_l = Ag_{kl}+Bu_ku_l$, with constant $A$ and $B$. The 
condition implies $R_{jkl}{}^m u_m = AB(u_j g_{kl}-u_kg_{jl})$,
which has the form \eqref{KDD}. Therefore, a concircular vector
is both Riemann and Weyl compatible.  
\end{remark}

The general statements valid for compatible tensors \cite{RCT} become
stronger for compatible vectors, and new facts arise. For example, 
the generalized Derdzi\'nski and Shen's theorem has now a surprisingly 
simple proof, with no need of auxiliary $K-$tensors:   
\begin{thrm}\label{34}
Let $K$ be a generalized curvature tensor, and $u$ be a $K$-compatible 
vector.\\ 
1) If $u^2\neq 0$ and $v$, $w$ are vectors orthogonal to 
$u$ ($u_av^a=0$, $u_a w^a=0$), then:
\begin{align}
K_{abcd}w^av^bu^c =0.
\end{align} 
2) If $u^2=0$ and $v$ is orthogonal to $u$ ($u_av^a=0$) then: 
\begin{align}
K_{abcd}u^av^bu^c =0.
\end{align}
\begin{proof}
1) The $K$-compatibility condition, 
$(u_a K_{bcde} + u_b K_{cade} + u_c K_{abde}) u^e =0 $,
is contracted with $u^av^bw^c$: 
$$(u^au_a) v^bw^cK_{bcde}u^e + u^a(u_bv^b)w^c K_{cade}u^e + u^av^b(w^cu_c) 
K_{abde} u^e =0. $$ 
The last two terms cancel because of orthogonality. 
2) The $K$-compatibility condition is contracted with $u^av^b$ and equation
$u_cK_{abde}u^av^bu^e =0$ is obtained. The result follows if $u$ is non zero.
\end{proof}
\end{thrm}

\begin{remark}
For the Riemann tensor, $R_{abcd}v^aw^bu^c $ is
the vector obtained through parallel transport of $u$ along a 
parallelogram with infinitesimal vectors $v$ and $w$. It is known that,
if  $R_{abcd}v^aw^bu^c =0$ for any $v$ and $w$, then it is
$R_{abcd}u^c =0$. If $u$ is Riemann compatible, then it has zero 
variations along infinitesimal parallelograms with directions orthogonal 
to it. 
\end{remark}

\begin{thrm}
Let $X(1), \ldots ,X(n)$ be an orthonormal basis of a $n$-dimensional 
pseudo-Riemannian manifold, $X(a)_kX(b)^k=\pm \delta_{ab}$. If $X(3)\ldots
X(n)$ are Riemann compatible, then all Pontryagin forms vanish.
\begin{proof}
Among three vectors, one is certainly Riemann compatible; therefore
it is always $R_{ij}{}^{kl}X(a)^i\wedge X(b)^j X(c)_k =0$, by Theorem \ref{34}. 
This means that the column vectors of the matrix
$R_{ij}{}^{kl}X(a)^i\wedge X(b)^j$ are orthogonal to all vectors
$X(c)$ with $c\neq a,b$, i.e. they belong to the subspace spanned by $X(a)$ 
and $X(b)$. Because of the antisymmetry in $k,l$, it is necessarily
$$R_{ij}{}^{kl}X(a)^i\wedge X(b)^j = \lambda_{ab} X(a)^k\wedge X(b)^l$$ 
This condition of 
pureness of the Riemann tensor implies the vanishing of all
Pontryagin forms (\cite{RCT} Theor.~5.2).
\end{proof}
\end{thrm}

The identity \eqref{RC} relating Riemann and Weyl compatibility, 
is rewritten for vectors:
\begin{align}\label{kkC}
(u_a C_{bclm} + u_b C_{calm} + u_c C_{ablm})u^m
= (u_a R_{bclm} + u_b R_{calm} + u_c R_{ablm})u^m\\
+ \frac{1}{n-2}\left [
g_{cl} u_{[a} R_{b]m}  + g_{al} u_{[b} R_{c]m}  + g_{bl} u_{[c} R_{a]m} 
\right ]u^m. \nonumber
\end{align}
A first consequence is the restatement of 
theorem \ref{bRWRicci} for vectors:
\begin{prop}\label{uRCRicci}
A vector field $u$ is Riemann compatible if and only if it is Weyl compatible 
and $u_{[a} R_{b]}{}^mu_m =0$. 
\end{prop}
\noindent
On a Lorentzian manifold the proposition is equivalent to Prop.~4.2  
(time-like vectors) and Prop.~A13 (null vectors) in Ref.~\cite{Ortaggio2}.

A second consequence is the extension of a theorem by Hall, which he proved 
for null vectors in $n=4$ space-times \cite{Hall}. It is valid 
in any dimension  and metric signature, and for vectors not necessarily null: 
\begin{thrm}
Consider the following conditions on a vector field $u$: 
\begin{align*}
A)\quad  &u_{[a}R_{b]clm} u^cu^m + u^2 R_{ablm}u^m =0,\\
B)\quad  &u_{[a} C_{b]clm}u^cu^m + u^2 C_{ablm}u^m=0,\\
C)\quad  &u_{[a} R_{b]m}u^m =0.
\end{align*}
Two conditions imply the third one. In particular, if $u^2\neq 0$ the
stronger statement holds: $A$ is true if and only if $B$ and $C$ are true.
\begin{proof}
Eq.\eqref{kkC} is contracted with $u^c$,
\begin{align}
 u_{[a} C_{b]clm}u^cu^m +u^2 C_{ablm}u^m
= u_{[a} R_{b]clm}u^cu^m +u^2R_{ablm}u^m \label{HALL}\\
 + \frac{1}{n-2}\left [ u_lu_{[a} R_{b]m}u^m + (g_{al} u_b  - 
g_{bl} u_a)u^cu^m R_{cm} -u^2 (g_{al}R_{bm}-g_{bl}R_{am})u^m\right ].  
\nonumber
\end{align}
If condition $C$ is true, its contraction with $u^b$ gives 
$(u_au^b R_{bm} -u^2 R_{am})u^m =0$ and \eqref{HALL} becomes
$ u_{[a} C_{b]clm}u^cu^m + u^2C_{ablm}k^m = 
u_{[a} R_{b]clm}u^cu^m +u^2 R_{ablm}u^m $.
Therefore $B$ and $C$ imply $A$, or $A$ and $C$ imply $B$.\\
Suppose now that $A$ is true; contraction of condition $A$ by $g^{al}$
gives $u^2 R_{bm}u^m - u_b(u^cu^m R_{cm})=0$, and \eqref{HALL} becomes: 
\begin{align}
 u_{[a} C_{b]clm}u^cu^m +u^2 C_{ablm}u^m
=  \frac{1}{n-2} u_lu_{[a} R_{b]m}u^m \label{99}
\end{align}
Validity of $A$ and $B$ imply that $u_lu_{[a} R_{b]m}u^m=0$ i.e.
$C$ is true.\\
A stronger result holds if $u^2\neq 0$. Contraction of \eqref{99} by 
$u^l$ makes the left-hand-side vanish and condition C is true. 
Then, the same equation \eqref{99} states that also B is true, i.e. 
A implies B and C.
\end{proof}
\end{thrm}

\begin{remark} Condition $C$ is met in Einstein spaces, defined by
$R_{ab}-\frac{1}{n}\,R\, g_{ab}=0$.
\end{remark}

\begin{remark} Condition $B$ plays a special role in the classification of 
manifolds. Some cases where it holds are:
1) $u^m R_{abcm}=0$ {\rm (\cite{McIvLee,Stephani2})}; 
2) $k$ is a recurrent null vector, $\nabla_a k_b=\lambda_ak_b$, 
with $\nabla_{[a}\lambda_{b]}=0$ {\rm (\cite{Stephani} page 69)}; 
3) Manifolds with constant curvature {\rm (\cite{Stephani} page 101)}:
$$ R_{bclm} = \frac{R}{n(n-1)} (g_{bl} g_{cm}-g_{cl}g_{bm}) $$
In cases 1,2 the vector is Riemann compatible.
\begin{proof}
1) The relation implies $R_{am}u^m=0$. Then the whole r.h.s. of \eqref{kkC}
is zero and $(u_aC_{bclm}+u_bC_{calm}+u_c C_{ablm})u^m=0$. Multiply by $u^c$
and obtain $A$.\\ 
2) $[\nabla_a,\nabla_b]u_c= R_{abc}{}^mu_m$; because of recurrency and 
closedness, the l.h.s. is $\nabla_a(\lambda_b u_c)-\nabla_b(\lambda_a u_c)=
(\lambda_b\nabla_a-\lambda_a\nabla_b)u_c =0$. Then case 1) is obtained.\\
3) Contraction with $g^{cm}$ shows that the manifold is Einstein, then 
condition $C$ holds. If $u$ is a vector, obtain $u_a R_{bclm}u^cu^m = 
\frac{R}{n(n-1)}u_a(g_{bl}u^2-u_bu_l)$; then $u_{[a,} R_{b]clm}u^m= 
u^2 \frac{R}{n(n-1)}(u_ag_{bl}-u_bg_{al}) = -u^2R_{ablm}u^m$ i.e. condition
$B$ is true, and $B$ and $C$ imply $A$. 
\end{proof}
\end{remark}
%
%
%

\section{Permutable vectors}
In the same way that compatibility is defined for vectors,
permutability of a vector is defined by permutability of the tensor
$u_iu_j$:  
\begin{definition}\label{vectpermut}
A vector is Riemann (Weyl) permutable if $R_{kl[i}{}^mu_{j]}u_m=0$
($C_{kl[i}{}^mu_{j]}u_m=0$). 
\end{definition}
\noindent
On Lorentzian manifolds and for null vectors, the definition is 
equivalent to the Bel-Debever condition for Weyl type II(abd) (\cite{Ortaggio}
Table~1).

\begin{remark}\label{vp}
If $u$ is Riemann (Weyl) permutable and $u^2\neq 0$, then 
$R_{kljm}u^m=0$  ($C_{kljm}u^m=0$).
\end{remark}
A special class of Riemann permutable vectors is:
\begin{align}
R_{abc}{}^mu_m=0. \label{Ru}
\end{align}
Null vectors of this sort describe gravitational waves in Einstein's 
linearized theory (see \cite{Stephani2} page 244). 
A complete classification of space times that satisfy \eqref{Ru} 
is given in Theorem 1.1 of ref.\cite{McIvLee}.\\
Eq.\eqref{Ru} arises as 
the integrability condition for the equation 
$\nabla_a u_b+\nabla_bu_a = 2\lambda g_{ab}$ with constant $\lambda $
and the constraint  $\nabla_a u_b-\nabla_b u_a=0$
($u$ is a {\em homothetic} vector, see \cite{Stephani} pp.~69,~564).
Vectors that fullfill \eqref{Ru} also arise in the symmetric solution of 
the equation
\begin{equation}
R_{abc}{}^m x_{dm}+R_{abd}{}^m x_{cm}=0 \label{21}
\end{equation}
which, by the Ricci identity, is
equivalent to $[\nabla_a,\nabla_b ]x_{cd} = 0$.
It has a trivial solution $x_{ab}= \phi\, g_{ab}$, where $\phi$ is a scalar.
McIntosh and Halford \cite{McIHa} investigated spacetimes 
whose Riemann tensor admits a non trivial solution,
such as Einstein spaces, G\"odel metric, Bertotti-Robinson metric. 
McIntosh and Hall proved \cite{McIH} that the only nontrivial  solution is 
$x_{ab}=\alpha u_au_b$, where $u$ has the property \eqref{Ru} and 
$\alpha $ is a scalar field.\\
Besides the uniqueness stated above in $n=4$, we prove in general:
\begin{prop}
Let $x_{ab}$ be a symmetric tensor that fulfills \eqref{21}.\\
1) $x_{ab}$ is Riemann (and Weyl) compatible;\\
2) If $X$ and $Y$ are two eigenvectors, $X^m x_{cm} =\lambda X_c$ 
and $Y^m x_{cm} = \mu Y_c$ with $\lambda \neq \mu$, then 
$ R_{abcm} X^cY^m = 0$.
\begin{proof}
Summation on cyclic permutations of indices $abc$ in \eqref{21} gives a 
vanishing term (first Bianchi identity) and Riemann compatibility. Property
2 is proven exactly as in Prop.~\ref{permXY}.
\end{proof}
\end{prop}

\section{Petrov types and Weyl compatible vectors}
In 1954 Petrov classified $n=4$ space-times according to the 
degeneracy of the eigenvalues of the self-dual part of the Weyl tensor. 
The eigenvalues solve an equation of degree four \cite{Petrov}. 
In type I spaces they are distinct, in type II spaces two are coincident 
and two are distint, in type D spaces they are pairwise coincident, 
type III spaces have three equal eigenvalues, and finally 
in type N spaces all eigenvalues coincide \cite{Stephani2}. Type O spaces
are conformally flat. The same types arise in the classification by Bel 
and Debever 
\cite{Bel,Deb}, which is based on null vectors that solve increasingly
restricted equations:
\begin{align}
&\text{type}\,I    &k_{[b}C_{a]rs[q}k_{n]}k^rk^s=0\\
&\text{type}\, II,D     &k_{[b}C_{a]rsq}k^rk^s=0 \label{IID}\\
&\text{type}\,III       &k_{[b}C_{a]rsq}k^r=0 \label{III}\\
&\text{type}\,N         &C_{arsq}k^r=0\\
&\text{type}\,O         &C_{arsq}=0
\end{align}
When at least two vectors $k$ are degenerate, i.e. $k$ meets condition 
\eqref{IID}, the Weyl tensor is named {\em algebraically special} 
\cite{Sachs,Stephani2}.\\
The classification was generalized to $n>4$ and includes the above
relations \cite{Coley2008,Coley,Ortaggio2,Ortaggio}.\\ 

Let's consider the above classification in the perspective of
Weyl compatibility. According to the general definition  
\eqref{uKcomp}, a vector is Weyl compatible if
\begin{align}
(u_i C_{jklm} + u_j C_{kilm} + u_k C_{ijlm}) u^m =0.\label{uCcomp}
\end{align} 
\begin{thrm}\label{nullkalgspec}
On a Lorentzian manifold, if a null vector $k$ is Weyl 
compatible (or Riemann compatible), then the Weyl tensor is algebraically 
special.
\begin{proof}
Multiply \eqref{uCcomp} by $k^c$ and use the antisymmetry of Weyl's 
tensor:\\
$0=(k_a C_{bcd}{}^m + k_b C_{cad}{}^m)k^ck_m= k_{[a}C_{b]cd}{}^m k^ck_m
= - k_{[a}C_{b]cmd} k^ck^m $.
\end{proof}
\end{thrm}
\noindent
In ref.\cite{Ortaggio} (Table~1), the condition \eqref{uCcomp} 
is the statement that the Weyl tensor is type II(d).\\
If a space-time admits a null concircular vector,
$\nabla_k u_l= Ag_{kl}+Bu_ku_l$, then the Weyl tensor is algebraically 
special (see Remark \ref{concircular}).\\
For a null-dust $n$-dimensional space-time, $T_{ab}=\Phi^2\,k_ak_b$ (eq.5.8
in \cite{Stephani}), the condition $D\Pi_l=0$ is verified if and only if
the Weyl tensor is type II(d) with respect to $k$ (see theorem \ref{DPi}).\\ 
The theorem extends Theorem 1.1 in \cite{McIvLee}, which holds
for null vectors such that $R_{ijk}{}^mk_m=0$.\\

Space-times with a null Weyl-compatible vector are Petrov type II 
or D. Are they more special than II or D? In general the answer is no.
In a type III space-time there are three coincident principal 
directions, i.e. there is a null vector such that $k_{[b}C_{a]rsq}k^r=0$. 
This means that the null $k$ is Weyl-permutable, a property
that implies Weyl compatibility (see def. \ref{vectpermut}):
\begin{prop}
A null vector $k$ solves \eqref{III}, which corresponds to $n=4$ 
space-times of Petrov type III, if and only if it is Weyl-permutable.\\
\end{prop}


\begin{prop} 
If $u_k u_l$ is a Codazzi tensor and $u_j$ is a closed 1-form, then:\\
1) if $u^2=0$ the Weyl tensor is algebraically special;\\
2) if $u^2\neq 0$ the integral curves of $u$ are geodesic lines.
\begin{proof}
1) Codazzi tensors are Riemann compatibile, and thus Weyl compatible. If 
moreover $u^2=0$, prop. \ref{nullkalgspec} applies. 
2) The Codazzi condition $\nabla_a (u_b u_c) = \nabla_b(u_a u_c )$ 
and closedness $\nabla_au_b=\nabla_b u_a$ give
$u_b \nabla_a u_c  - u_a \nabla_b u_c = 0$. Exchange $a$ with $c$ and 
subtract to obtain: $u_c\nabla_b u_a - u_a\nabla_b u_c = 0$. Multiply
by $u^cu^b$: 
\begin{align*}
 (u^b \nabla_b )u_a =   \left[\frac{u^cu^b\nabla_b u_c}{u^2}\right]\, u_a
\end{align*} 
i.e. the integral curves of $u$ are geodesics 
(see \cite{DeFelice}, eqs. 2.9.4 and 2.9.5).
\end{proof}
\end{prop}

\section{Electric and magnetic tensors}
In a $n=4$ space-time the Weyl tensor has 10 independent 
components that can be accounted for by two symmetric tensors. Given a 
vector $u$ with $u^au_a=-1$, the electric and magnetic components of the 
Weyl tensor are \cite{Berts}:
\begin{align}
E_{ab} = u^ju^m C_{jabm} \,  ,\qquad  H_{ab} = u^ju^m \tilde C_{jabm} 
\end{align}
where $\tilde C_{abcd}=\frac{1}{2}\epsilon_{abrs}C^{rs}{}_{cd}$ 
is the dual tensor. The two tensors are symmetric, traceless, and
satisfy $E_{ab}u^b=0$, $H_{ab} u^b=0$. Then they each have 5 independent
components, and completely describe the Weyl tensor.\\
If the electric and magnetic components are proportional, $\nu E=\mu H$ for
some scalar fields $\mu $ and $\nu $ (including the case when one of them is 
zero), the space is type I, D or O \cite{Stephani} (page 73). The following
theorem was partly proven in \cite{RCT} and is stated in \cite{Ortaggio} for 
any $n$:
\begin{thrm}
On a $n=4$ space-time, a vector $u$ is Weyl-compatible if and only if $H=0$.
\begin{proof}
Consider the following chain of identities:
\begin{align}
H^a{}_b &=u_ju^m \tilde C^{ja}{}_{bm} = 
\frac{1}{2}u_ju^m C_{rsbm}\epsilon^{jars}
\nonumber \\
&=\frac{1}{6}[u_jC_{rsbm}\epsilon^{jars} + u_rC_{sjbm}\epsilon^{rasj}+
u_sC_{jrbm}\epsilon^{sajr}]u^m  \nonumber \\
&=\frac{1}{6}[u_jC_{rsbm} + u_rC_{sjbm}+
u_sC_{jrbm}]u^m \epsilon^{jars} \label{HWeyl}
\end{align}
The equality shows that $H=0$ is equivalent to Weyl compatibility.
\end{proof}
\end{thrm}
It follows that a $n=4$ space-time with a Weyl compatible time-like
vector is type I, D or O. This extends Theorem 1.1 in \cite{McIvLee}.\\ 
If a spacetime admits a time-like concircular vector $\nabla_ku_l = Ag_{kl}
+Bu_ku_l$, with constant $A$ and $B$, then the magnetic part vanishes. 

\begin{thrm}
A $n=4$ space-time with a non-null Weyl 
permutable vector is conformally flat, $C_{jkl}{}^m=0$ (type O).
\begin{proof}
Let $E$ and $H$ be the electric and magnetic components evaluated with 
$u$. If $u$ is Weyl permutable, then it is Weyl compatible and $H=0$. Let's
show that also $E$ is zero. Multiply the relation \eqref{uCcomp} for
Weyl compatibility by $u^j$:
$u^2 C_{kilm}u^m= u_k E_{il}-u_i E_{kl}$.
Because $u$ is Weyl permutable, it is $C_{kilm}u^m=0$
(see remark \ref{vp}); then 
$0= u_k E_{il}-u_i E_{kl}$. Multiply by $u^k$ and use $u^kE_{kl}=0$ to
obtain $E_{il}=0$. 
\end{proof} 
\end{thrm}

The definitions of electric and magnetic components of the Weyl tensor 
can be generalized 
by replacing the symmetric tensor $u^iu^j$ by a symmetric tensor $T^{ij}$:
\begin{align}
E_{ab} = T^{jm}C_{jabm}, \qquad H_{ab} = T^{jm}\tilde C_{jabm}. 
\end{align}
\begin{prop} $E$ and $H$ are symmetric and traceless.
\begin{proof} 
The first statement follows from the symmetry of $C_{ijkl}$ or 
$\tilde C_{ijkl}$ in the exchange of $ij$ with $kl$, and symmetry of $T$. 
The second follows from tracelessness of the Weyl tensor and its dual. 
\end{proof}
\end{prop}

\begin{prop}{\quad}\\
1) If $T$ is Weyl-compatible then $E$ commutes with $T$;\\ 
2) $H=0$ if and only if $T$ is Weyl compatible. 
\begin{proof} 
The proof is based on the following identities: 
\begin{align}
E_{ab}T^b{}_c-T_a{}^bE_{bc} = 
&-[T_{cb}C_{jam}{}^b+T_{ab}C_{cjm}{}^b+T_{jb}C_{acm}{}^b]T^{jm}\\
H^a{}_b =&\frac{1}{6}[T_j{}^m C_{rsbm} + T_r{}^mC_{sjbm}+
T_s{}^mC_{jrbm}] \epsilon^{jars}
\end{align}
The second identity is proven along the same line as \eqref{HWeyl}.
The first equation is proven here:
\begin{align*}
E_{ab}T^b{}_c-T_a{}^bE_{bc} =& [C_{jabm}T^b{}_c-C_{jbcm}T_a{}^b]T^{jm}\\
=&-[T^b{}_cC_{jamb}+T^b{}_aC_{cjmb}+T^b{}_j C_{acmb}]T^{jm}\nonumber\\
=&-[T_{cb}C_{jam}{}^b+T_{ab}C_{cjm}{}^b+T_{jb}C_{acm}{}^b]T^{jm}
\end{align*}
where the last term added in the second line is identically zero. 
\end{proof}
\end{prop}


\section{Hypersurfaces}
Let $\mathscr M_n$ be a hypersurface in a pseudo Riemannian manifold 
$(\mathbb V_{n+1},\tilde g)$. The metric tensor (first fundamental form) is 
$g_{hj}=\tilde g(B_h,B_j)$, where $B_1\ldots B_n$ are the tangent vectors.
If $N$ is the vector normal to the hypersurface it is 
$\tilde g (B_h, N) =0$. The Riemann tensor 
is given by the Gauss equation \cite{Lovelock}: 
\begin{align} 
R_{jklm} = \tilde R_{\mu\nu\rho\sigma}
B^\mu{}_jB^\nu{}_k B^\rho{}_l B^\sigma {}_m 
\pm\, (\Omega_{jl}\Omega_{km} -\Omega_{jm}\Omega_{kl} )
\label{Gauss}
\end{align}
with a symmetric tensor $\Omega_{ij}$ (second fundamental form)
constrained by the Codazzi equation:
\begin{align}
\nabla_k \Omega_{jl} - \nabla_j \Omega_{kl} = N^\mu 
\tilde R_{\nu\mu\rho\sigma}
B^{\nu}{}_j B^\rho{}_l B^\sigma{}_k \label{Codazzi}
\end{align}
If $\mathbb V_{n+1}$ is a constant curvature manifold, the Gauss and Codazzi 
equations simplify:
\begin{align} 
&R_{jklm} =
\frac{\tilde R}{n(n+1)} (g_{jl}g_{km} -g_{jm}g_{kl} )
\pm\, (\Omega_{jl}\Omega_{km} -\Omega_{jm}\Omega_{kl} ),
\label{Gauss2}\\
&\nabla_k \Omega_{jl} - \nabla_j \Omega_{kl} = 0, \label{Codazzi2}
\end{align}
If  $\mathbb V_{n+1}$ is (pseudo)-Euclidean, the terms proportional 
to the scalar curvature $\tilde R$ vanish  \cite{Stephani}. For this case
Stephani proved that eqs. \eqref{Gauss2} and \eqref{Codazzi2} are sufficient 
conditions 
for a manifold $\mathscr M_4$ to have an embedding in $\mathbb V_5$ 
\cite{Stephani} (page 587). A general theorem by Goenner, 
restricted to pseudo-Euclidean manifolds $\mathbb V_{n+1}$, states that 
if $\Omega_{kl}$ is 
invertible, then it is a Codazzi tensor \cite{Stephani} (page 587).
A simple proof of the same fact is here given, for the case
of constant curvature $\mathbb V_{n+1}$:
\begin{thrm} 
If $R_{jklm}$ has the form \eqref{Gauss2} and $\Omega $ is invertible, 
then $\Omega $ is a Codazzi tensor.
\begin{proof}
The second Bianchi identity for the Riemann tensor is 
\begin{align*}
\Omega_{mk} (\nabla_i \Omega_{jl} - \nabla_j \Omega_{li} ) + 
\Omega_{mi} (\nabla_j \Omega_{kl} - \nabla_k \Omega_{lj}) + 
\Omega_{mj} (\nabla_k \Omega_{il} - \nabla_i \Omega_{lk} ) +\\
+\Omega_{jl} (\nabla_i \Omega_{mk} - \nabla_k \Omega_{im}) + \Omega_{kl} 
(\nabla_j \Omega_{mi} - \nabla_i \Omega_{jm} ) + \Omega_{il} 
(\nabla_ k \Omega_{mj} - \nabla_j \Omega_{km} ) = 0
\end{align*}
Moltiplication by $(\Omega^{-1})^{km}$ gives: 
\begin{align*}
&(n-3) (\nabla_i \Omega_{jl} - \nabla_j \Omega_{li} ) \\
&= 
-(\Omega^{-1})^{km}[\Omega_{jl}(\nabla_i \Omega_{mk} - \nabla_k \Omega_{im}) + 
\Omega_{il}(\nabla_ k \Omega_{mj} - \nabla_j \Omega_{km} )]. 
\end{align*}
Multiplication by $(\Omega^{-1})^{lj}$ gives: 
$2(n-2) (\Omega^{-1})^{lj}(\nabla_i \Omega_{jl} - \nabla_j \Omega_{li} )=0$. 
This result is used to simplify the previous equation:
$(n-3) (\nabla_i \Omega_{jl} - \nabla_j \Omega_{li} )=0$, which for $n>3$ 
is the Codazzi property.
\end{proof}
\end{thrm}




\begin{thrm}
Let $\mathscr M_n$ be a hypersurface isometrically embedded in a 
pseudo-Riemannian space $\mathbb V_{n+1}$ with constant curvature. Then:\\
1) $\Omega $ is Weyl compatible;\\
2) the eigenvectors of $\Omega $ are Weyl compatible;\\
3) the Ricci tensor is Weyl compatible.
\begin{proof}
1) For a hypersurface that is isometrically embedded in a constant
curvature space, $\Omega_{ij}$ is a Codazzi tensor, and then it is both
Riemann and Weyl compatible.\\  
2) Given the form \eqref{Gauss2} of the Riemann tensor, if 
$\Omega_{km}u^m= \lambda u_k$ then:
$$ u_i u^m R_{jklm} = k u_i(u_k g_{jl} - u_j g_{kl} )
\pm \lambda u_i(\Omega_{jl}u_k - u_j \Omega_{kl})  $$
where, for shortness, $k=\tilde R/n(n+1)$.
Summation on cyclic permutations of $i,j,k$ cancels all terms in the 
right-hand-side, and one is left with Riemann compatibility:
$u_i u^m R_{jklm} + u_j u^m R_{kilm} + u_k u^m R_{ijlm} = 0$.\\
3) The Ricci tensor for a hypersurface isometrically embedded in a
costant curvature space is $R_{kl} = \pm (\Omega^2_{kl} - 
\Omega_p{}^p\Omega_{kl}) + k (n-1) g_{kl}$. 
Let us first show that $\Omega^2$ is Riemann compatible. Evaluate the 
expression $(\Omega^2)_{im} R_{jkl}{}^m + 
(\Omega^2)_{jm} R_{kil}{}^m + (\Omega^2)_{km} R_{ijl}{}^m $ with the
Riemann tensor \eqref{Gauss2}. The first term is:
\begin{align*}
\Omega^2_{im} R_{jkl}{}^m = k 
\left ( g_{jl} \Omega^2_{ik} -  g_{kl} \Omega^2_{ij}\right)
\pm\left (\Omega_{jl} \Omega^3_{ik} - \Omega_{kl} 
\Omega^3_{ij}\right)
\end{align*}
While summing on cyclic permutations of $ijk$, all terms in  the r.h.s cancel. 
Therefore the tensor $\Omega^2$ is Riemann compatible and thus Weyl 
compatible. 
Since the Ricci tensor is the sum of Riemann - compatible terms, it 
is itself Riemann compatible, and thus Weyl compatible.
\end{proof}
\end{thrm}
By considering Einstein's equation \eqref{EE} one also has:
\begin{cor}
On a space-time that is isometrically embedded as a 
hypersurface is a pseudo Riemannian space $\mathbb V_{n+1}$ with constant
curvature, the energy momentum tensor is Weyl compatible.\\
If the energy momentum 
tensor has the form $T_{kl} = a u_k u_l + b g_{kl}$ with $u^i u_i = -1$, then 
the Weyl tensor is purely electric.
\end{cor}

\section{Geodesic maps}
Let $(\mathscr M,g)$ be a pseudo-Riemannian manifold. A geodesic map 
$\mathscr M\to
\mathscr M$ induces a pseudo-Riemannian structure $(\mathscr M,\tilde g)$ with 
Christoffel symbols 
$\tilde\Gamma^k_{ij} = \Gamma^k_{ij} + \delta_i{}^k X_j + 
\delta_j{}^k X_i$, where $X$ is a closed 1-form \cite{Formella,Mikes,Sinyukov}.
Accordingly, the new Riemann tensor is
$\tilde R_{jkl}{}^m = R_{jkl}{}^m +\delta_j{}^m P_{kl} -\delta_k{}^m P_{jl}$,
with {\em deformation tensor} $P_{kl} = \nabla_k X_l - X_k X_l$. Since $X$ is
closed, the deformation tensor is symmetric. The new Ricci tensor 
is $\tilde R_{kl} =R_{kl} - (n-1)P_{kl}$.

In \cite{RCT} we showed that for geodesic maps the following identity holds
for any symmetric tensor:
\begin{align}
b_{im} \tilde R_{jkl}{}^m + b_{jm} \tilde R_{kil}{}^m + 
b_{km} \tilde R_{ijl}{}^m = 
b_{im} R_{jkl}{}^m + b_{jm} R_{kil}{}^m + b_{km} R_{ijl}{}^m;\label{btildeRbR}
\end{align}
as a consequence, the property of Riemann compatibility is conserved.
What about Weyl compatibility? For general symmetric tensors the answer 
is difficult by the fact that the expression \eqref{Weyltensor} for the 
new Weyl tensor contains $\tilde g$, which is not
simply related to $g$. This is a sufficient condition:

\begin{prop}
If a symmetric tensor $b$ commutes with the Ricci and the
deformation tensors, then
\begin{align}
b_{im} \tilde C_{jkl}{}^m + b_{jm} \tilde C_{kil}{}^m + 
b_{km} \tilde C_{ijl}{}^m = 
b_{im} C_{jkl}{}^m + b_{jm} C_{kil}{}^m + b_{km} C_{ijl}{}^m
\label{btildeCbC}
\end{align}
\begin{proof}
If $b$ commutes with the Ricci and the deformation tensors, then it commutes
with $\tilde R_{ij}$. With these conditions, \eqref{RC} implies that
\begin{align*}
b_{im} C_{jkl}{}^m + b_{jm} C_{kil}{}^m + b_{km} C_{ijl}{}^m 
= b_{im} R_{jkl}{}^m + b_{jm} R_{kil}{}^m + b_{km} R_{ijl}{}^m 
\end{align*}
and the same relation with tensors $\tilde C_{jkl}{}^m$ and 
$\tilde R_{jkl}{}^m$. Since \eqref{btildeRbR} holds for geodesic maps, 
\eqref{btildeCbC} follows.
\end{proof}
\end{prop}
\noindent
A simplification occurs for {\em special geodesic maps}, defined by the 
property  $P_{kl} = \gamma \, g_{kl}$, meaning that $X$ 
is a concircular vector: 
$\nabla_k X_l - X_k X_l = \gamma\, g_{kl}$ \cite{DeszczHotlos}.

\end{document}